\documentclass[preprint,pre,aps]{revtex4}
\usepackage[dvips]{graphicx}
\usepackage{graphics}
\usepackage{amsmath,amssymb}
\bibstyle{prsty}
\begin{document}
\title{Origin of similarity of  phase diagrams  in  amphiphilic  and colloidal systems with 
competing interactions}
\author{ A. Ciach, J. P\c ekalski and W. T. G\' o\' zd\' z}
\affiliation{Institute of Physical Chemistry,
 Polish Academy of Sciences, 01-224 Warszawa, Poland}
 \date{\today} 

\begin{abstract}
We show that  amphiphilic and colloidal  systems with competing interactions can be described by the same
 Landau-Brazovskii functional. 
The functional is obtained by a systematic coarse-graining procedure applied to
systems with isotropic interaction potentials.
Microscopic expressions for the coefficients in the functional are derived. 
We propose simple criteria to distinguish the effective
 interparticle potentials that can lead to macro- or microsegregation.
 Our considerations concern also charged globular proteins in aqueous solutions and other
 system with effective short-range attraction long-range repulsion interactions. 
\end{abstract}
\maketitle 
\section{Introduction}
 Recent experimental \cite{stradner:04:0,campbell:05:0,klix:10:0,elmasri:12:0}, simulation
 \cite{candia:06:0,imperio:06:0,archer:07:1} 
and theoretical studies
 \cite{tarzia:06:0,ortix:08:0,archer:08:0,ciach:08:1,ciach:10:1,roth:11:0} have revealed striking
 similarity between colloidal and amphiphilic self-assembly, 
despite 
different interaction potentials in such systems. Interactions between amphiphilic molecules are strongly 
orientation-dependent, whereas
effective interactions between spherical colloid particles usually depend only on the distance between
 their centers. 
When particles are charged and polymers are present in solution,
 short-range depletion attraction competes with long-range electrostatic repulsion
 \cite{dijkstra:99:0,stradner:04:0,campbell:05:0,elmasri:12:0} (SALR potential).
 The SALR potential is important for many other soft-matter and biological systems,
 because both the colloid particles and the
 macromolecules in water are typically charged and repell each other with screened electrostatic forces,
 and in addition attract each other with van der Waals and 
 solvent-induced solvophobic\cite{israel:11:0,shukla:08:0,iglesias:12:0} or thermodynamic
 Casimir forces \cite{veatch:07:0,hertlein:08:0,gambassi:09:0,machta:12:0}.
 Attraction leads 
to cluster formation, but further growth of the clusters is suppressed by sufficiently strong repulsion
~\cite{mossa:04:0}  when the
 size of the cluster becomes comparable with the range of the repulsion. 
For increasing concentration of colloidal particles spherical clusters (droplets), elongated clusters, slabs, 
cylindrical voids (bubbles)  and 
spherical voids  were seen in  MC simulation\cite{archer:07:1}. When the clusters, rods, slabs 
 or voids are periodically
 distributed in space, ordered soft crystals are formed. The
hexagonal and lamellar  phases and transitions
 between them
were discovered in MD \cite{candia:06:0} 
and MC \cite{imperio:06:0} simulations. Transitions between the lamellar phase and the hexagonal phases of droplets
and bubbles were also predicted by density functional theory (DFT)~\cite{archer:08:0}.
 For intermediate densities  a  network of particles was obtained by Brownian dynamics  simulations
\cite{sciortino:05:0,toledano:09:0}.
Similar phase diagrams but with micelles, reverse micelles, bilayers and  bicontinuous phases instead of, respectively,
 clusters (droplets), voids (bubbles), slabs and  networks of particles
 were obtained for water-surfactant 
mixtures and for 
block copolymers\cite{gompper:94:3,leibler:80:0,matsen:96:0,seul:95:0}  (see Fig.1).

  Experimental studies confirmed existence of spherical and elongated clusters, as well as particle  networks
 \cite{stradner:04:0,campbell:05:0,klix:10:0,elmasri:12:0}. The ordered phases have not been discovered yet, 
because many long-lived metastable states are present, and in
confocal microscopy observations 
\cite{campbell:05:0,elmasri:12:0} instantaneous, 
rather than average distributions of particles were observed. Average distribution can be obtained from dynamical 
confocal microscopy measurements, and in the future such experiments should be conducted in order to obtain phase 
diagrams. Reliable determination of phase diagrams in simulations is also nontrivial because of the 
long-lived metastable states \cite{toledano:09:0}, and
suitable  simulation procedures (e.g. the one suggested in Ref.~\cite{wilding:13:1})
 are required. 
Because of these difficulties, theoretical predictions are necessary to  guide both experiment and simulations.

The purpose of this work is to understand origin of the similarity between the phase diagrams in amphiphilic systems
and in systems interacting with the SALR type potentials. Once the mapping between the mathematical models
describing these systems is established, one can take advantage of the results obtained earlier for the 
amphiphilic systems in studies of systems with competing interactions.

\begin{figure}
\includegraphics[scale=0.62]{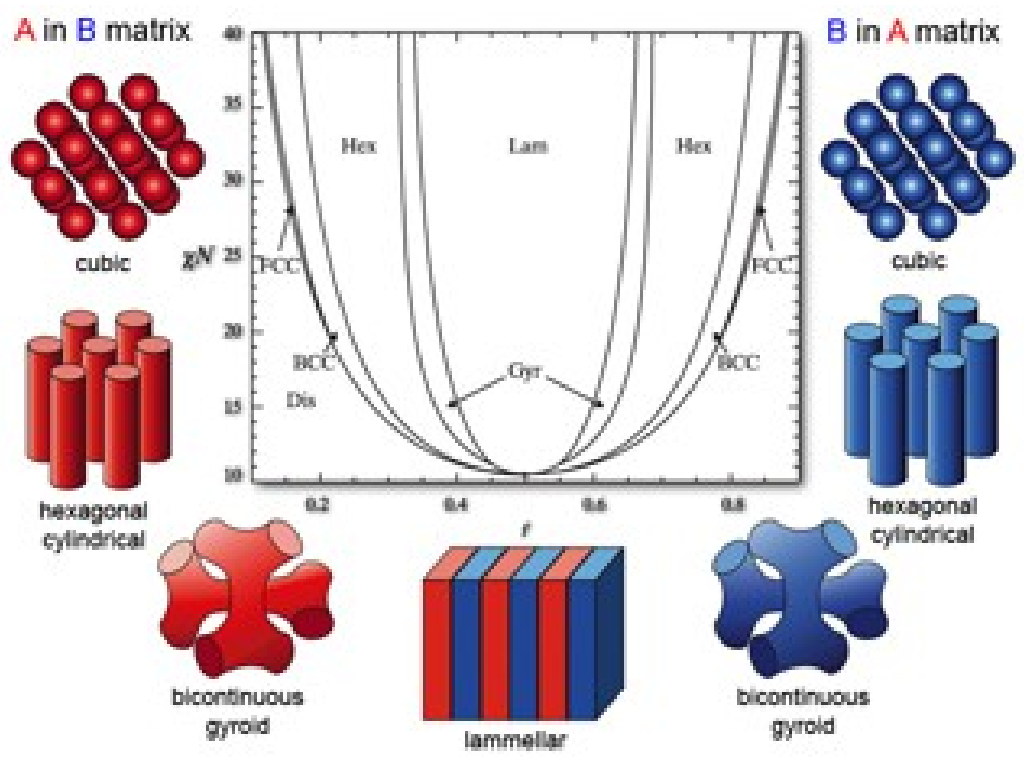}
\includegraphics[scale=0.92]{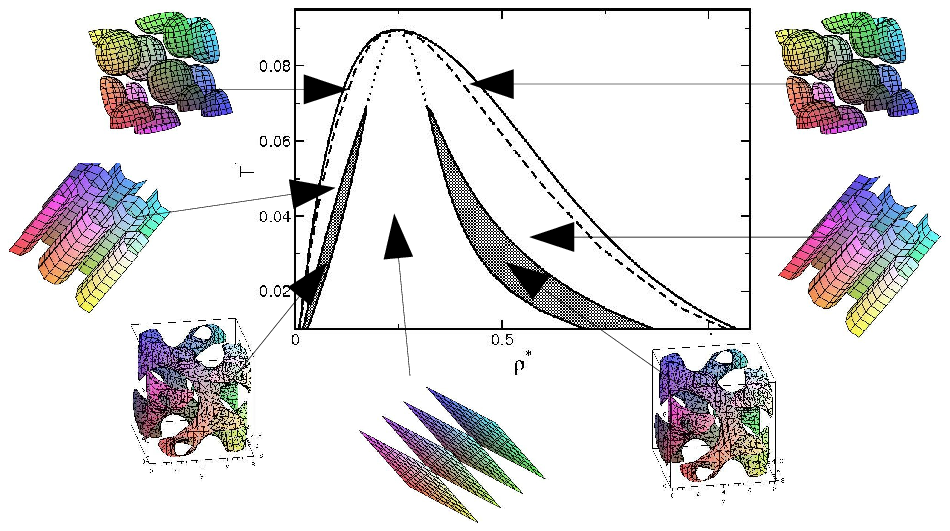}
\caption{Top panel: phase diagram in block copolymer system (reprinted form Ref.\cite{matsen:96:0}).  $f$ is the ratio
 between the number of the A and B monomers and the Flory parameter $\chi$ is inversely proportional to
 temperature $T$.
 Bottom panel: phase diagram in the SALR-potential system obtained in Refs.\cite{ciach:08:1,ciach:10:1}. 
$\rho^*=6\eta/\pi$,
 where $\eta$ is the  volume fraction of particles. The structure of the phases stable in regions separated by
 the solid, dashed and dotted lines is illustrated by the corresponding surfaces placed arround the phase diagram.
 Inside the regions enclosed by these surfaces the density is enhanced or
 depleted compared to $\rho^*$ when $\rho^*<0.25$ or
 $\rho^*>0.25$ respectively.
}
\end{figure}

 It is well known that the  topology  of the phase diagrams
in  systems  
undergoing separation into homogeneous phases is the same  \cite{barrat:03:0}. 
This universality
 is reflected in the generic Landau functional \cite{barrat:03:0}
of the order parameter (OP) $\phi$  (e.g. a deviation of the density  from its critical value),
\begin{eqnarray}
\label{landau}
{\cal L}=\int \!\! d{\bf r} \,\Big[
f(\phi({\bf r})) + \frac{\beta V_2}{2}|\nabla \phi({\bf r})|^2
\Big],
\end{eqnarray}
where
$f(\phi)=(A_2/2+\beta V_0)\phi^2 + A_4\phi^4/4!$,
 $\beta=(k_BT)^{-1}$,  $k_B$ is the Boltzmann constant and $T$ is the temperature,  
$V_0<0$ is the measure of attraction, $A_n>0$ and $V_2>0$.
 The term $\beta V_2|\nabla \phi({\bf r})|^2/2$  ensures that the  phases corresponding to the minimum of 
the functional (\ref{landau}) are homogeneous ($\nabla \phi({\bf r})=0$).

In order to describe microsegregation in  systems with competing interactions, several authors 
 extended the functional (\ref{landau}) by including different repulsive terms 
\cite{andelman:87:0,tarzia:06:0,ortix:08:0}. 
Unfortunately, such an approach  suffers from an inconsistent treatment of the attractive and repulsive parts 
of interactions; while the former is included in the coarse-grained functional, the latter has a microscopic form.  
Universal emergence of modulated phases  was noted in these and other studies, but it remains unclear  why 
 the pattern formation in the presence of frustration should be the same as in amphiphilic systems.
 In contrast to  simple systems that are all described by the   same  functional (\ref{landau}), 
 universal ordering on the mesoscopic length scale was not related to a
 generic functional common for all microsegregating systems.

The functional  successfully used
 for block copolymers  and microemulsions has the 
form~\cite{leibler:80:0,fredrickson:87:0,seul:95:0,gompper:94:3,ciach:01:2}
\begin{eqnarray}
\label{braz}
{\cal L}_B=\int \!\! d{\bf r} \,\Big[
 f(\phi({\bf r}))+ \frac{\beta V_2}{2}|\nabla \phi({\bf r})|^2+ \frac{\beta V_4}{4!}\Big(\nabla^2 \phi({\bf r})\Big)^2
\Big]
\end{eqnarray}
with $V_2<0$  and $V_4>0$. The inhomogeneous structure is  favored and disfavored by the second and the third
 term in (\ref{braz}) respectively. 
Competition between these terms leads to a finite  length  scale of inhomogeneities,  $2\pi/k_b$, with
 $k_b^2=-6 V_2/V_4$~\cite{brazovskii:75:0}. We should mention that the functional (\ref{braz}) has essentially the same form as
 the free energy functional in the phase-field-crystal (PFC) model of freezing and pattern formation on atomistic 
length scale \cite{emmerich:12:0}. 

Because the functional (\ref{braz}) describes succesfully various inhomogeneus systems, is  plausible that  
the  generic model for  
systems with competing interactions has the same form, with $\phi({\bf r})$ denoting local excess volume fraction
 of particles. However, it is not obvious apriori if the functional 
(\ref{landau}) with $V_2>0$  or (\ref{braz}) with $V_2<0$ is appropriate
for a given form of interactions. Thus, it is necessary to find the relation between the coefficients
 in the functional and the form of the interaction potential. Such a relation can be reliably obtained when
 the functional (\ref{braz}) is derived from a microscopic theory.

In this work we consider   effectively one-component systems of particles interacting with  
spherically symmetric potentials of  arbitrary form. Solvent molecules and depletion agents are taken 
into account indirectly in the form of effective interactions between the particles. 
In sec.2 we derive approximate expression for the internal energy. 
In sec.3 we derive and discuss the Landau-Brazovskii functional.
 The effective potentials are classified in sec.4. Sec.5 contains short summary and discussion.

\section{Approximate expressions for the internal energy}

Let us focus on the internal energy (configurational part), 
\begin{eqnarray}
\label{U}
 {\cal U}=\frac{1}{2}\int \!\! d{\bf r} \!\! \int \!\! d\Delta{\bf r} \,\,
\rho({\bf r})\rho({\bf r}+\Delta{\bf r})
 V(\Delta r)g(\Delta r)
\end{eqnarray}
where $V(\Delta r)$ and  $g(\Delta r)$ 
are the interaction potential and the pair distribution
 function for particles located at ${\bf r}$ and ${\bf r}+\Delta{\bf r}$, 
and $\rho({\bf r})$ is the local average
 density of  the particles. We focus on systems inhomogeneous on a mesoscopic
 length scale and on weak ordering, 
therefore we assume that $g$
depends only on $\Delta r$, and 
 $\rho({\bf r})$ is a slowly varying function.  
\subsection{Short-range interaction potentials}

 In the first step we consider short-range interaction potentials, whose moments $\int d{\bf r} V(r)r^{n}$ 
are finite at least for $n=4$.  We Taylor expand $\rho({\bf r}+\Delta{\bf r})$ 
 about ${\bf r}$, integrate by parts  (see Appendix) and  obtain the following approximate expression 
for the internal 
energy (\ref{U})
\begin{eqnarray}
\label{U1}
  {\cal U}\approx \int \!\! d{\bf r} \,\big[V_0\eta({\bf
r})^2+\frac{V_2}{2}
|\nabla\eta({\bf r})|^2+...
\big]
\end{eqnarray}
where  $\eta({\bf r})=\rho({\bf r})v $ is the local volume fraction, $v=\pi\sigma^3/6 $ is
  the particle volume, and 
\begin{eqnarray}
 \label{Vnn}
V_n=\frac{2\pi (-1)^{n/2}}{(n+1)v^2}\int_{0}^{\infty} \!\!\! d r \,
r^{2+n} V( r)g(r).
\end{eqnarray}
For  attractive 
interactions ($Vg<0$)
 homogeneous phases  are energetically favored, because 
$V_2>0$, and the second term in (\ref{U1})  leads to an \textit{increase} of ${\cal U}$ for
 $\nabla\eta({\bf r})\ne 0$.
 Note that $g>0$ and either oscillates around 1 in  crystals or exhibits oscillatory decay to 1 in liquids. Thus, repulsion
 at large distances ($Vg>0$) can lead to $V_2<0$, and hence to a \textit{decrease} of ${\cal U}$ for 
$\nabla\eta({\bf r})\ne 0$, i.e. to spatial inhomogeneities.  
For an  illustration we plot $V(r)r^n$  in Fig.2  for the 
 double Yukawa  potential \cite{archer:07:1,archer:07:0}
\begin{equation}
\label{Yukawa}
 V(r)=-K_1\frac{\exp(-z_1r)}{r}+K_2\frac{\exp(-z_2r)}{r}.
\end{equation}
\begin{figure}
\includegraphics[scale=0.24]{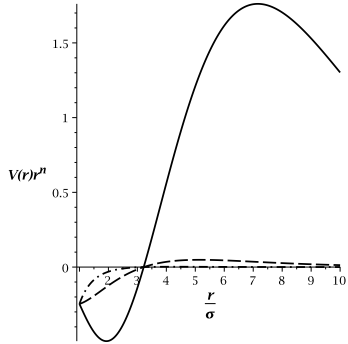}
\caption{$ V(r)r^n$  for $n=4$ (solid line), $n=2$ (dash line) and $n=0$
(dash-dot line), for the SALR potential (\ref{Yukawa}) with $ K_1=1, K_2=0.2,z_1=1,z_2=0.5$.
 Note that the large positive integrand can lead to a positive integral
$\int_0^{\infty}dr r^4V(r)$. Because  $g>0$ and for large distances approaches 1 
(or oscillates around 1 in the case of crystalline order inside the clusters), we obtain
 $V_2<0$  (see (\ref{Vnn})). Note also that the above argument is not 
restricted to the
potential (\ref{Yukawa}), but holds for any spherically-symmetric interaction potential assuming large positive value
for large $r$.
}
\end{figure}

For $V_2<0$ the Taylor expansion of 
$\rho({\bf r}+\Delta{\bf r})$ should be truncated at the fourth order term, and 
(\ref{U1}) should be replaced by (see Appendix)
\begin{eqnarray}
\label{U2}
 {\cal U}\approx \int \!\! d{\bf r} \,\Big[V_0\eta({\bf r})^2+\frac{V_2}{2}
|\nabla\eta({\bf r})|^2+\frac{V_4}{4!}\big(\nabla^2\eta({\bf r})\big)^2
\Big] .
\end{eqnarray}
From (\ref{Vnn}) it follows that  $V_4>0$ if $V_2<0$, and the above functional is stable. 
 Note the similarity between the last two terms in Eqs.(\ref{U2}) and (\ref{braz}). 
Spatial inhomogeneities favored by Eq.(\ref{U2}) for $V_2<0$ are consistent with preferential formation of clusters in 
the case of 
the SALR potential, with the  size and the distance between the clusters  determined by the range of 
attraction and the range of repulsion respectively.
\subsection{Long-range interaction potentials}
The above considerations are valid  for potentials that decay faster than $ 1/r^7$ 
(see (\ref{Vnn})).
For the long-range potentials whose moments (\ref{Vnn}) diverge, the expansion (\ref{U2}) of the internal 
energy  (\ref{U})
is not valid. 
However, the internal energy can be approximated by Eq.(\ref{U2}) even for long-range interactions, 
but with $V_n$ 
given in terms of the Fourier transform
\begin{equation}
\label{Vco}
\tilde V_g(k)=\int d{\bf r}\exp(i{\bf k}\cdot {\bf r})\frac{V(r)g(r)}{v^2}, 
\end{equation}
which for $k>0$ may exist even when the moments in (\ref{Vnn}) diverge.
Eqs.(\ref{U}) and (\ref{U2}) in Fourier representation take the forms
\begin{eqnarray}
\label{omix}
{\cal U}=
\frac{1}{2}
\int \frac{d{\bf k}}{(2\pi)^3}\tilde V_g(k)|\tilde \eta({\bf k})|^2,
\end{eqnarray}
and
\begin{eqnarray}
\label{Uk0}
 {\cal U}\approx \int\frac{d{\bf k}}{(2\pi)^3} \Big(
V_0+\frac{V_2}{2}k^2 +\frac{V_4}{4!}k^4
\Big)|\tilde \eta({\bf k})|^2 ,
\end{eqnarray}
where  $\tilde \eta( {\bf k})=\int \! d {\bf r}\exp(i{\bf k}\cdot{\bf r})\eta({\bf r})$.
  Note that Eq.(\ref{Uk0}) is obtained when  $\tilde V_g(k)$ in Eq.(\ref{omix})  is approximated by
 a truncated Taylor expansion in small $k$. When  $V_n$ defined in Eq.(\ref{Vnn})  diverge,
 $\tilde V_g(k)$ is nonanalytic at $k=0$ and 
Eq.(\ref{Uk0}) is not valid.

 In order to develop appropriate approximate expression for $U$ in such a case, 
let us consider a change of the internal energy (\ref{U}) per unit volume associated with 
formation of the mesoscopic inhomogeneity $\eta_0\to\eta({\bf r})=\eta_0+\phi({\bf r})$. For the planar
density-wave
$\phi({\bf r})=\phi_k\cos(kz)$ we have
\begin{eqnarray}
\label{Du}
 \Delta u=\phi_k^2\tilde V_g(k)/4.
\end{eqnarray}
When the amplitude of the density modulations $\phi_k$ is fixed, $\Delta u$ takes the minimum at $k=k_b$,
 corresponding to the minimum of $\tilde V_g(k)$. Thus, the energetically favored length scale of 
inhomogeneity is  $2\pi/k_b$.
Spatial inhomogeneities can lead to a decrease of the internal energy if
\begin{equation}
\label{crit}
\tilde V_{g}(k_b)<0.
\end{equation}
If $k_b>0$ and is not very small, then $\tilde V_g(k)$ should be expanded about its minimum at $k=k_b$, 
rather than about $k=0$ where it may be nonanalytic.
 When the minimum of $\tilde V_g(k)$ at  $k=k_b$ is deep, the waves with  $k\approx k_b$ 
 lead to much lower $\Delta u$ (see (\ref{Du}))  than the waves with $k$ significantly different from $k_b$.
 For this reason the waves with  $k\approx k_b$ 
are thermally excited with much higher
 probability than the waves with $k$ significantly different from $k_b$.
 For the potentials with deep minimum  at  $k=k_b\ne 0$  the expansion of $\tilde V_g(k)$ about $k=k_b$ 
can be truncated, because for small values of $|k-k_b|$ the
truncated Taylor expansion is close to $\tilde V_g(k)$. On the other hand,
 the waves with the wavenumbers $k$  much different from $k_b$ 
 are excited with negligible probability and can be disregarded.
If we require that the approximate expression for $\tilde V_g$ is an even function of $k$, then for 
$ k_b>0$ we obtain 
\begin{equation}
\label{Vga}
 \tilde V_g ( k)\approx \tilde V_g ( k_b)+ v_2(k^2-k_b^2)^2/2
\end{equation}
 where $v_2=\tilde V_g^{''}(k_b)/(2k_b)^2$.  
When (\ref{Vga}) is inserted in (\ref{omix}) and  the relations between 
 $k^n\tilde \eta({\bf k})$ and $\nabla^n \eta({\bf r})$ are taken into account, then Eq.(\ref{omix})
 in real space 
representation takes  the form 
 (\ref{U2}) but with $V_n$ given by
\begin{equation}
\label{VF}
 V_0=\frac{1}{2}(\tilde V_g(k_b)+v_2k_b^4/2),V_2=-v_2k_b^2,V_4=6v_2.
\end{equation} 
The above holds only for $k_b>0$.
 We finally stress that Eq.(\ref{U2}) can be used
 provided that (\ref{Vga}) is a reasonable approximation for $ \tilde V_g(k)$; if it is not, the internal energy
  given in Eq.(\ref{omix})
 cannot be  simplified according to the above scheme and the functional (\ref{U2}) is not valid.
 We further discuss this issue in sec.4.

\section{Derivation of the Landau-Brazovskii functional}

 In order to  find  thermal equilibrium we need to compare grand potentials
 in systems with and without mesoscopic inhomogeneities.
A particular form of the volume fraction on the mesoscopic scale, $\eta({\bf r})$,
imposes a constraint on the volume occupied by the particles  in mesoscopic regions \cite{ciach:08:1}.
Let us consider the grand potential in the presence of the constraint $\eta({\bf r})$,
\begin{equation}
\label{omco}
 \Omega_{co}[\eta]=U[\eta]-TS[\eta]-\bar\mu \int d{\bf r}\eta({\bf r}),
\end{equation}
where 
$U[\eta]$ and  $S[\eta]$ are
 the configurational parts of 
the internal energy and the entropy,
 $\bar\mu=[ \mu-k_BT\ln(\Lambda/\sigma)^3]/v$ and  $\Lambda$ is the thermal wavelength.
We assume that $U[\eta]$ is given by (\ref{U1}) or (\ref{U2}), except that $g$ in (\ref{Vnn}) and (\ref{Vco}) 
 should be replaced by $g_{co}$ calculated  
for the fixed mesoscopic state $\eta({\bf r})$.

When $\eta({\bf r})$  varies on a length scale larger than $\sigma$, 
then we can make the local density approximation
for the  entropy,
 $-TS\approx\int \!\! d{\bf r} \, [f_h(\eta({\bf r}))]$,
where $f_h(\eta)$ is
 the configurational part of 
the free energy density of the hard-sphere reference system
 with the volume fraction $\eta$.  
$f_h(\eta_0+\phi({\bf r}))$ can be Taylor expanded.
 Note that  by definition the local 
volume fraction $\eta({\bf r})=v\rho({\bf r})<1$.
Local deviations $\phi({\bf r})=\eta({\bf r})-\eta_0$  
from the space-average volume fraction $\eta_0$ are 
 $|\phi({\bf r})|< 1$,  and the truncation of the Taylor series for $f_h(\eta_0+\phi({\bf r}))$
 is justified. For weak ordering
 ($\phi({\bf r})\ll 1$) $f_h(\eta)$
can be approximated by the  polynomial in $\phi$.

From  (\ref{omco}), (\ref{U1}), (\ref{U2}) and the above we can see that the change of $\beta \Omega_{co}$ 
associated with  creation of the mesoscopic inhomogeneity,
\begin{equation}
\label{lzet}
{\cal L}_{\eta_0}[\phi]=\beta\Omega_{co}[\eta_0+\phi]-\beta\Omega_{co}[\eta_0],
\end{equation}  
 takes the form of the 
functional (\ref{landau}) or  
(\ref{braz}) for $V_2>0$ or  $V_2<0$ respectively,  
 with 
\begin{equation}
\label{f}
 f(\phi)=\sum_{n\ge 1}\frac{1}{n!}\frac {d^n \beta f_h(\eta)}{d\eta^n}|_{\eta=\eta_0}\phi^n+2\eta_0 \beta V_0\phi
 + \beta V_0\phi^2-\beta\bar\mu\phi.
\end{equation}

Thus, we have shown that self-assembly in  amphiphilic systems and in
 systems with competing interactions  can be described by the same
functional (\ref{braz}). 
This explains  striking similarity of the phase diagrams. 

The phase diagrams  in Fig.1
and earlier results \cite{leibler:80:0,gompper:94:3} obtained by minimization of the functional (\ref{braz}) 
are in good  qualitative agreement with the DFT results \cite{archer:08:0},
 and  both theories are in qualitative 
agreement with simulations. In simulations the modulated phases are stable in a
 smaller region of the phase diagram (the disordered phase becomes stable at lower $T$ than found in MF),
 and the transition
 between the disordered and lamellar phases occurs for a much larger temperature 
interval \cite{candia:06:0,imperio:06:0}.  
To explain these discrepancies, let us note that  
 $\Omega_{co}[\eta]$ is calculated for the specified mesoscopic volume fraction $\eta({\bf r})$. A
probability of  spontaneous appearance  of $\eta({\bf r})$
  is $ p[\eta]\propto \exp(-\beta\Omega_{co}[\eta])$ \cite{ciach:08:1,barrat:03:0}. 
The inhomogeneous and homogeneous states occur with the same probability
 when  ${\cal L}_{\eta_0}[\phi]=0$, 
since $p[\eta_0+\phi]/p[\eta_0]= 
\exp(-{\cal L}_{\eta_0}[\phi])$.
This should not be mistaken with thermodynamic 
equilibrium associated with equality of the grand potentials 
\begin{equation}
\label{Omega}
 \beta\Omega[\bar\eta]=\beta\Omega_{co}[\bar\eta]-\ln \Big[ \int \!\! 
D\psi \, e^{-\beta H_{fluc}[\bar\eta,\psi]}\Big]
\end{equation}
where $\bar\eta({\bf r})$ is the \textit{average} mesoscopic volume fraction at which $\Omega[\bar\eta]$ 
takes the global
 minimum, and $ H_{fluc}[\bar\eta,\psi]=\Omega_{co}[\bar\eta+\psi]-\Omega_{co}[\bar\eta]$ with
 $\psi({\bf r})$ denoting the
 mesoscopic fluctuation (displacement of denser regions with respect to their average positions)
 \cite{ciach:08:1}. 
When the  fluctuation contribution  in Eq.(\ref{Omega})
is included as in Ref.\cite{brazovskii:75:0}, then the modulated phases are stable
 for a smaller region of the phase diagram. In addition, the transition between the disordered 
and lamellar phases occurs for 
a larger temperature range~\cite{fredrickson:87:0,podneks:96:0}, in better agreement with simulations 
\cite{candia:06:0,imperio:06:0}. Mesoscopic fluctuations dominate at the high-$T$ part of the phase diagram,
 therefore for high $T$ this approach is more accurate than mean-field theories. 
Due to our assumptions ($\phi\ll 1$, $k_b\sigma\ll \pi$),  for large values of the OP (i.e. for low $T$) 
 the microscopic DFT is superior.

Note that Eq.(\ref{Omega}) allows for further improvement of the theory by combined DFT and field-theoretic
 approaches applied to the first and the second term respectively. The first term in (\ref{Omega}) can be
 improved by assuming a more accurate DFT expression for $\Omega_{co}[\bar\eta]$.
On the other hand, the dominant fluctuations in the second term in (\ref{Omega}) correspond to small 
values of $ \beta H_{fluc}[\bar\eta,\psi]$, since the Boltzmann factor $\exp(-a)$ takes negligible values 
for large $a$. Thus, the dominant fluctuation contribution comes from small  $\psi$, for which 
$\Omega_{co}[\bar\eta+\psi]-\Omega_{co}[\bar\eta]$ can be Taylor expanded, and the field-theoretic methods
 can be used for evaluation of the second term in (\ref{Omega}).  Preliminary steps in this direction are 
described in Ref.\cite{ciach:08:1}.

\section{Classification of the interaction potentials}

Homogeneous systems can become unstable with respect to the density wave with the wavelength $2\pi/k_b$,
 when the condition (\ref{crit}) is satisfied. In the case of the density wave of the form $\phi_k\cos(k_bz)$ 
a region of excess density, 
$\pi/k_b$, is followed by a region of depleted density of the same size. In the case of weak ordering 
to which this theory is restricted,  
the density wave does not deviate much from superposition of plane waves  in different directions.
Effective potentials satisfying (\ref{crit}) can be classified as attraction-dominated ($k_b\approx 0$), 
repulsion dominated ($k_b\sigma\simeq \pi$) 
or competing ( $0\ll k_b\sigma\ll \pi$) (Fig.3). The attraction dominated potentials 
can lead to gas-liquid transition, since the region  with excess  density, $\pi/k_b\to \infty$, is macroscopic. 
The repulsion
 dominated potentials can lead to periodic ordering of individual particles, since  
the region with excess density, $\pi/k_b\sim \sigma$ (followed by the region of depleted density of the same size),
 is comparable with the size of the  particles. 
The competing potentials can lead to 
excess density in mesoscopic regions, $\sigma \ll \pi/k_b\ll\infty $.

The numerical values of $V_n$ and $k_b$ depend  on the shape of the interaction potential and on the
 approximation for the pair distribution function $g_{co}(r)$, calculated
under the constraint of fixed volume fractions in mesosocpic regions, $\eta({\bf r})$.  $g_{co}(r)$   
is considered as an imput from a microscopic theory to our mesoscopic 
description  and should obey $g_{co}(r)\to 1$ for $r\to\infty$ and $g_{co}(r)\to 0$ for $r\to 0$. 
Let us compare the results for 
 $g_{co}=1$ as in simple local DFT theories, and $ g_{co}(r)=\theta(r/\sigma-1)$ (unit step function), for which 
the contributions to $U[\eta]$ 
from overlapping hard spheres are not included. 

For the double Yukawa potential (Eq.(\ref{Yukawa})) we have
\begin{eqnarray}
\label{vk0}
 \tilde V(k)=4\pi\Bigg[\frac{K_2}{z_2^2+k^2}-\frac{K_1}{z_1^2+k^2}\Bigg]
\end{eqnarray}
for $g_{co}=1$, whereas for $g_{co}=\theta(r-1)$ (we set $\sigma\equiv 1$)
\begin{eqnarray}
\label{vk}
 \tilde V_g(k)=4\pi\Bigg[\frac{K_2e^{-z_2}}{z_2^2+k^2}
\Big(z_2\frac{\sin k}{k}+\cos k\Big)-\frac{K_1e^{-z_1}}{z_1^2+k^2}
\Big(z_1\frac{\sin k}{k}+\cos k\Big)\Bigg].
\end{eqnarray}
Eqs. (\ref{vk}) and (\ref{vk0}) are shown for the same parameters in Figs.3 and 4 respectively.
 Note that 
except from the pure repulsion, both approximations give similar positions $k_b$ of the global minimum,
  and similar shape 
 of $ \tilde V_g(k)$ for $k\approx k_b$, but the value at the minimum is different.

At the crossover between the gas-liquid separation and the periodic ordering  the minimum 
of $\tilde V_g(k)$ at $k=0$ 
becomes a maximum (Fig.3). Near the crossover the  minimum of $\tilde  V_g(k)$ is shallow 
 (i.e. $\tilde V_g^{''}(k_b)\to 0$). Eq.(\ref{Vga}) is valid when the minimum of 
$\tilde  V_g(k)$ at $k=k_b$ is deep, therefore it  is an oversimplification 
 near the crossover between the gas-liquid separation and the periodic ordering.

In order to easily verify if a given potential leads to microsegregation, we note that
at the crossover between gas-liquid separation and periodic ordering  the second derivative of $\tilde V_g(k)$ 
at $k=0$ vanishes. For the double Yukawa potential (Eq.(\ref{Yukawa})) 
this happens
 for  $K_2/K_1=K_{cr}$,  where $K_{cr}$ depends on the approximation for $g_{co}$. 
The micro-separation or the gas-liquid transition can occur when 
 $K_2/K_1\gg K_{cr}$ or $K_2/K_1<K_{cr}$ respectively.
When $g$ is neglected as in Eq.(\ref{vk0}), we obtain 
\begin{equation}
K_{cr}=(z_2/z_1)^4,
\end{equation}
whereas for $g_{co}=\theta(r-1)$ from (\ref{vk}) we obtain
\begin{equation}
\label{Kratio}
K_{cr}=\Big(\frac{z_2}{z_1}\Big)^4\frac{\Big(2+2z_1+z_1^2+
\frac{z_1^3}{3}\Big)e^{z_2}}{\Big(2+2z_2+z_2^2+\frac{z_2^3}{3}\Big)e^{z_1}}.
\end{equation}

For $z_1=1$ and $z_2=0.5$  we have $K_{cr} = 1/16$, or  $K_{cr}\approx 0.061$ for $\tilde V_g (k)$ given 
in Eq. (\ref{vk0})
or (\ref{vk}), in semiquantitative agreement with $K_{cr}\approx 0.059$ and $K_{cr}\approx 0.053$ in the
self-consistent Ornstein-Zernicke approximation (SCOZA) and nonlocal DFT  (see Ref.\cite{archer:07:0}). 
For increasing $z_1$ 
the discrepancy between $K_{cr}$ obtained for the
two different forms of $g_{co}$ increases. Nevertheless,
 \begin{equation}
\label{crY}
 K_2/K_1\gg (z_2/z_1)^4 
 \end{equation}
can serve as a simple condition 
that must be satisfied by potentials (\ref{Yukawa}) that can lead to microsegregation. 
Similar criterion was obtained in Refs.\cite{tarzia:06:0,mossa:04:0}. 
In Ref. \cite{tarzia:06:0} the maximum of the structure factor for $k>0$ was required for the microsegregation, 
in analogy to our approach. In Ref. \cite{mossa:04:0} it was assumed that the microsegregation may occur if 
 the energy per particle takes a minimum in a finite cluster.

The structure factor $S(k)$ is more accurate for  $g_{co}=\theta(r-1)$  than 
for $g_{co}=1$, because in the first case
the contributions to the internal energy from overlapping cores
of the particles are not included.
The structure factor was obtained in Refs.\cite{ciach:08:1,ciach:10:1} in an approximation analogous to
 the random phase approximation (RPA), 
\begin{equation}
 S(k)=\frac{\tilde G(k)}{\rho},
\end{equation}
where
\begin{equation}
\label{RPA}
\tilde G(k)^{-1}=\frac{v^2 \delta^2\beta\Omega_{co}}{\delta\tilde\eta({\bf k})\delta\tilde\eta(-{\bf k})}
=v^2\Bigg(\beta\tilde V_g(k)+
\frac{d^2\beta f_h}{d\eta^2}\Bigg).
\end{equation}
$S(k)$ takes a maximum at $k=k_b>0$ for both $g_{co}=\theta(r-1)$  and $g_{co}=1$  when ordering on 
the mesoscopic length scale occurs.  Another maximum at
$k\sigma\approx 2\pi$,  resulting from packing of hard spheres, is present for $g_{co}=\theta(r-1)$ 
and absent for
$g_{co}=1$. This second maximum of $\tilde G(k)$ results from the minimum of $\tilde V_g(k)$ at
 $k\sigma\approx 2\pi$ (Fig.3). 

Further studies are necessary to find the best approximation for $g_{co}$, and to determine the effect of
 ordering 
on the microscopic length scale (described by  $g_{co}$) on the values of $V_n$.
However, our main conclusion that $V_2$ may be negative when the repulsion  is sufficiently strong at large
 distances ($r\gg\sigma$) remains valid for the pair distribution function $g_{co}>0$ oscillating around 
1 on the microscopic length scale, as in clusters exhibiting internal crystal-like order.
 The difference between the distribution  of the ordered and disordered clusters on the
 mesoscopic length scale, described by the functional
 (\ref{braz}), can result from different numerical values of the parameters $V_n$ 
which are influenced by the form of $g_{co}$.
\begin{figure}
\includegraphics[scale=0.35]{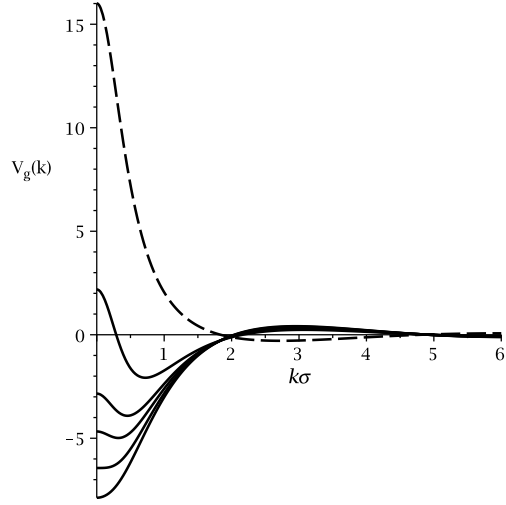}
\caption{ $\tilde  V_g(k)$ (Eq.(\ref{Vco})) for the  potential
(\ref{Yukawa}) and $g=\theta(r/\sigma-1)$.
 Dashed line: pure repulsion with $K_1=0, K_2/v^2=0.35,z_2=0.5$.
 Solid lines:  $z_1=1$, $z_2=0.5$,
 and $K_1/v^2=1$. From the bottom to the top lines  $K_2/v^2=0.03,0.06,0.1,0.14,0.25$.
}
\end{figure}

\begin{figure}
\includegraphics[scale=0.35]{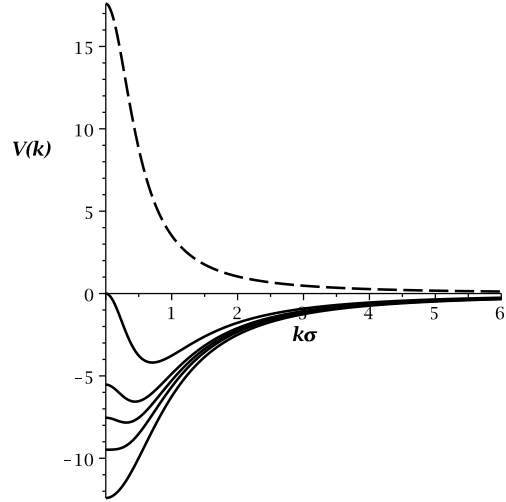}
\caption{Fourier transform $\tilde  V(k)$ (Eq.(\ref{vk0})) of the double Yukawa  potential.
 Dashed line: pure repulsion with $K_1=0, K_2/v^2=0.35,z_2=0.5$.
 Solid lines:  $z_1=1$, $z_2=0.5$,
 and $K_1/v^2=1$. From the bottom to the top lines  $K_2/v^2=0.03,0.06,0.1,0.14,0.25$.
}
\end{figure}

\section{Summary   and discussion}
We have derived the functional (\ref{braz}) for systems with competing interactions. 
The same functional (\ref{braz}) was successfuly used for amphiphilic
 systems\cite{leibler:80:0,gompper:94:3,podneks:96:0,fredrickson:87:0,ciach:01:2}, and in the 
PFC model of ordering on the atomistic scale \cite{emmerich:12:0,teeffelen:09:0,archer:12:0}. Our result
 supports on mathematical grounds the hypothesis of universality of microsegregation.

The functional (\ref{braz}) was intensively investigated in the context of block copolymers and microemulsions, 
and one can take advantage of these earlier results for the SALR systems.
In particular, based on the very
low surface tension between water and microemulsion, we can expect very low surface tension between 
homogeneous and modulated phases.  From the stability or metastability of the bicontinuous  phases 
 obtained from the functional (\ref{braz})\cite{gompper:94:3,podneks:96:0,ciach:01:2,gozdz:96:1} we expect
thermodynamic stability or metastability  of a 
network of particles (gel) \cite{podneks:96:0,ciach:10:1}. In Refs.\cite{candia:06:0,toledano:09:0,charbonneau:07:0}
gel  formation was interpreted as arrested microphase segregation. However,
 stable or very long-lived networks were also found \cite{zaccarelli:07:0,sciortino:05:0,campbell:05:0,klix:10:0}. 
According to our theory, the stable disordered network should be analogous to a bicontinuous
 microemulsion or a sponge phase. In addition, an
 ordered network of particles (gyroid phase) 
should be thermodynamically stable
for a narrow range of thermodynamic variables  (Fig.1). In systems with competing interactions such an orderd  phase 
has not been detected experimentally yet. 
 One can expect that aging of gels  is influenced by
 the structure of the
 thermodynamically stable phase in given thermodynamic conditions.
For a narrow range of thermodynamic variables corresponding to stability of the gyroid phase (Fig.1)
the percolating structure of the gel should never be destroyed, and may even become more regular. 
This fact may have practical implications, 
and verification of our prediction in future experimental and simulation studies or dynamic 
theories such as PFC \cite{emmerich:12:0,teeffelen:09:0,archer:12:0}    is important. 
 
Our considerations concern spherically symmetric (effective) potentials which in Fourier representation have a 
well-defined
 minimum for $0\le k_b\sigma<\pi$, but otherwise are of arbitrary form. Thus, our conclusions  are valid for
 a wide class of 
systems. In contrast to previous phenomenological theories,  we have obtained microscopic expressions 
for all the coefficients
 (Eqs.(\ref{Vnn}) or (\ref{VF}) and (\ref{f})). These expressions allow to predict if for given interactions macro
or microsegregation, described by the functional (\ref{landau}) or (\ref{braz}) respectievly, should occur. 
For the particular case of the double Yukawa potential (\ref{Yukawa}) we find a simple criterion (\ref{crY}) 
for potentials that may lead to microsegregation.

Our functional  is very similar to the  PFC model, and it is interesting to highlight the similarities and
 differences between the two theories. The main difference concerns  the interpretation of the OP field. 
 In the PFC the OP is the relative 
local deviation of the density from the space-averaged value, and the PFC is supposed to describe structure 
formation on the atomistic level. 
We consider volume fraction of particles averaged over mesoscopic regions, $\eta({\bf r})$, 
and the OP is the local deviation $\phi({\bf r})$
of this quantity  from the space-average value. The $\eta({\bf r})$ describes the volume 
occupied by the particles  within the mesocopic region 
around ${\bf r}$, therefore it may correspond to different positions of the particles. 
Thus, $\eta({\bf r})$  contains less precise information than the microscopic density.
As a result of the 'smearing' the particles over mesoscopic regions, the amplitude and 
the gradient of $\eta({\bf r})$    are both smaller than in the case of the microscopic density.
 The gradient expansion of $\phi({\bf r})$ in our derivation is better justified than the
 analogous expansion in derivation of the PFC from the DFT. Another consequence of the mesoscopic
 OP is the separation of all fluctuations  into microscopic fluctuations  for fixed 
 $\eta({\bf r})$ (such as displacements of the particles inside the clusters 
for fixed distribution of the clusters) and the mesoscopic fluctuations  represented by 
 different forms of $\eta({\bf r})$ (such as displacements of the clusters as a whole). 
Destructive role of mesoscopic fluctuations for periodic ordering on the mesoscopic 
length scale can be taken into account within field-theoretic methods
 (see Eq.(\ref{Omega}))\cite{brazovskii:75:0,ciach:08:1}. On the other hand,
 in the atomistic-level PFC all fluctuations are considered on the same footing.
 Derivations of the two theories are based on somewhat different further  
approximations. We consider separately the internal energy and the entropy. 
For the latter we assume the hard-sphere form in the local density approximation.
 In derivation of the PFC from the DFT the free energy is split in the ideal
 ${\cal F}_{id}$ and the excess ${\cal F}_{ex}$ parts, and in ${\cal F}_{ex}$ 
terms beyond the quadratic part are neglected.  While in our theory the 
coefficients $A_n$ with $n\ge 3$ are determined by the hard-sphere free energy 
in the local density approximation, in the PFC they have the  ideal gas form.  
On the other hand,  in the PFC the accuracy of the part quadratic in the OP field is
 limited only by the approximation for the direct correlation function. 

On the formal level the direct correlation function in our theory is given by 
an approximation similar to the RPA
 (see Eq.(\ref{RPA})), plus the contribution from the mesoscopic fluctuations,
 obtained from the second term in  (\ref{Omega}). Let us comment that smaller 
amplitude of density waves in the PFC than in the DFT \cite{emmerich:12:0} could be
 explained by reinterpretation of the OP along similar lines as in our derivation \cite{ciach:08:1}. 
 On the other hand, dynamics in systems with competing interactions could be describd by a theory
 analogous to the dynamical PFC \cite{emmerich:12:0,teeffelen:09:0,archer:12:0}.

Let us finally note that our derivation is based on the  assumption that $\tilde V_g(k)$ has a single, 
well defined global minimum for $0\le k <\pi$. Physically relevant
interaction potentials studied in 
Refs.\cite{stradner:04:0,campbell:05:0,elmasri:12:0,candia:06:0,imperio:06:0,archer:07:0,archer:07:1,tarzia:06:0,ortix:08:0,shukla:08:0,archer:08:0,ciach:08:1,ciach:10:1,toledano:09:0} have this property. An exception is the crossover between 
gas-liquid separation and periodic ordering, where the minimum of  $\tilde V_g(k)$ is very shallow.
 However, in general there exist functions with two or more  minima with the same or comparable depths 
 for $0\le k <\pi$.
 Such forms of $\tilde V_g(k)$ would correspond to simultaneous ordering on different length scales. 
It is not clear if interactions
with such a form of $\tilde V_g(k)$ are physically relevant, and if the hierarchical self-assembly may be associated
 with 
effective interactions that in Fourier representation have the above mentioned property. Such systems, if exist, 
 cannot be described by the Brazovskii functional (Eq.(\ref{braz})). 

  We hope that  our predictions can stimulate  experimental and simulation studies in this
 important, but still largely 
unexplored field. We stress that the periodic order concerns the average density, and due to 
the presence of fluctuations and large time scales, proper data
analysis is required to detect the order on the mesoscopic length scale.

 \section{Acknowledgments}
The work of JP  was   realized within the International PhD Projects
Programme of the Foundation for Polish Science, cofinanced from
European Regional Development Fund within
Innovative Economy Operational Programme "Grants for innovation".
 AC and WTG acknowledge the financial support by the NCN grant 2012/05/B/ST3/03302
\section{Appendix}

We derive an approximate expression for the internal energy ${\cal U}$ when the local density
 (or volume fraction)  varies on a mesoscopic length scale, i.e. for small gradients of $\rho({\bf r})$. 
When $\rho({\bf r}+\Delta {\bf r})$ is approximated by a truncated Taylor series, Eq.(\ref{U}) 
 takes the approximate form
\begin{eqnarray}
 \label{Usi}
 {\cal U}=\frac{1}{2}\int \!\! d{\bf r} \!\! \int \!\! d\Delta{\bf r} \,\,V(\Delta r)g(\Delta r)\rho({\bf r})\Big[
\rho({\bf r})+\Delta r_i\frac{\partial\rho}{\partial  r_i}+
\frac{1}{2}\Delta r_i\frac{\partial^2\rho}{\partial r_i\partial  r_j}\Delta r_j+...
\Big]
\end{eqnarray}
where $ {\bf r}=(  r_1, r_2, r_3)$ 
and summation convention for repeated indexes is used. 
The above can be written in the form 
\begin{eqnarray}
\label{U2si}
  {\cal U}=\int \!\! d{\bf r} \Big[ V_0 v^2\rho({\bf r})^2- 
\frac{V_2v^2}{2}\rho({\bf r})\sum_{i=1}^3\frac{\partial^2\rho}{\partial  r_i^2} +...\Big] 
\end{eqnarray}
where $V_0v^2=\int \!\! d{\bf r} V(r)g(r)/2$ and $-V_2v^2=\int \!\! d{\bf r}r_i^2 V(r)g(r)/2$.
 In derivation of the above we took into account that 
an integral over $R^3$ of an odd function, $ V(r)g(r)r_i$ and $V(r)g(r)r_ir_j$ with $i\ne j$, vanishes. 
When the second term in (\ref{U2si}) is integrated by parts and  the  boundary term is neglected,
 we obtain for ${\cal U}$ Eq.(\ref{U2}). 
Next we take into account that the integration over the angles in spherical variables for any function 
$f(r)$ of $r=|{\bf r}|$
 gives
\begin{eqnarray}
 \int d{\bf r} f(r)=4\pi\int_0^{\infty} dr r^2f(r) 
 \end{eqnarray}
and
 \begin{eqnarray}
  \int d{\bf r} f(r) r_i^2=
\frac{4\pi}{3}\int_0^{\infty} dr r^4 f(r), 
 \end{eqnarray}
and for $V_0$ and $V_2$ defined above we obtain Eq.(\ref{Vnn}).

 When $V_2<0$, the  functional in Eq.(\ref{U2}) is unstable, and  the Taylor expansion in 
(\ref{Usi}) must be
 truncated at the fourth order term. The  term associated with the third-order derivative vanishes,
 because the integrand is
  an odd function. In order to evaluate the fourth-order term, we perform the integration
 over the angles in spherical variables of
the integrands of the form $V(r)g(r) r_i^2r_j^2$
 and integrate by parts twice the expressions 
\begin{eqnarray}
\label{U3si}
\int \!\! d{\bf r}\rho({\bf r})\frac{\partial^4\rho}{\partial  r_i^2\partial  r_j^2} 
\end{eqnarray}
with $i=j$ as well as  $i\ne j$. We neglect the boundary terms, and after some algebra we finally 
obtain Eqs.(\ref{U2})  and (\ref{Vnn}).

\end{document}